# Favorable Interfacial Chemomechanics Enables Stable Cycling of High Li-Content Li-In/Sn Anodes in Sulfide Electrolyte Based Solid-State Batteries

Christian Hänsel[a], Baltej Singh[b], David Kiwic[a], Pieremanuele Canepa[b,c,*], and Dipan Kundu[d,*]

[a] Multifunctional Materials, Department of Materials, ETH Zurich, Switzerland

[b] Department of Materials Science and Engineering, National University of Singapore, Singapore 117575, Singapore.

[c] Department of Chemical and Biomolecular Engineering, National University of Singapore, 4 Engineering Drive 4, Singapore 117585

[d] School of Chemical Engineering, UNSW Sydney, Kensington 2052, Australia

* Corresponding author. Email: pcanepa@nus.edu.sg

* Corresponding author. Email: d.kundu@unsw.edu.au

**ABSTRACT**

Solid-state batteries (SSBs) can offer a paradigm shift in battery safety and energy density. Yet, the promise hinges on the ability to integrate high-performance electrodes with state-of-the-art solid electrolytes. For example, lithium (Li) metal, the most energy-dense anode candidate, suffers from severe interfacial chemomechanical issues that lead to cell failure. Li alloys of In/Sn are attractive alternatives, but their exploration has mostly been limited to the low capacity (low Li content) and In rich $Li_xIn$ ($x \leq 0.5$). Here, the fundamental electro-chemo-mechanical behavior of Li-In and Li-Sn alloys of varied Li stoichiometries is unravelled in sulfide electrolyte based SSBs. The intermetallic electrodes developed through a controlled synthesis and fabrication technique display impressive (electro)chemical stability with $Li_6PS_5Cl$ as the solid electrolyte and maintain nearly perfect interfacial contact during the electrochemical Li insertion/deinsertion under an optimal stack pressure. Their intriguing variation in the Li migration barrier with composition and its influence on the observed Li cycling overpotential is revealed through combined computational and electrochemical studies. Stable interfacial chemomechanics of the alloys allow long-term dendrite free Li cycling (>1000 h) at relatively high current densities (1 mA cm$^{-2}$) and capacities (1 mAh cm$^{-2}$), as demonstrated for $Li_{13}In_3$ and $Li_{17}Sn_4$, which are more desirable from a capacity and cost consideration compared to the low Li content analogues. The presented understanding can guide the development of high-capacity Li-In/Sn alloy anodes for SSBs.

**INTRODUCTION**

Inorganic solid electrolytes (SEs) can potentially serve as a mechanical barrier against Li metal dendrites and thus enable the safe use of the lithium metal anode in solid state batteries (SSBs).[1, 2] However, propagation of lithium filaments through microcracks/pores and grain boundaries inside the SEs, especially at high currents have shown that even highly dense inorganic SEs cannot successfully prevent dendritic short-circuit.[3] The main problem is not the SE itself but rather the insufficient contact between the SE and the metallic lithium electrode, morphological instabilities of lithium during anodic dissolution and the limited chemical stability of most SEs in contact with lithium metal.[4, 5] These factors lead to an increase of the interfacial resistance and inhomogeneous current distributions that both initiate and accelerate dendrite formation and propagation through the SE.[6] Thus, the practical application of metallic lithium anodes also faces grave challenges in SBBs. Recent studies have shown that external pressure during cell operation can minimize the problem of interfacial contact by deformation induced creep of the lithium electrode.[7-9] However, correct dosing of the pressure becomes challenging due to the extrusion of Li through the SE pores, which accelerates the short-circuit cell failure as SE film gets thinner.[10, 11]

As a result, there has been a renewed interest in lithium alloys as alternatives to the metallic lithium anode in SSBs. In solid-state cells assembled with $Li_7La_3Zr_2O_{12}$ (LLZO), lithium alloys based on Ge, Al, Sn, Au, Si, Mg, and Ag have been predominantly used as interfacial coatings between the garnet electrolyte and the lithium electrode.[12] These thin lithophilic coatings are aimed at enhancing the surface wettability of Li metal on LLZO and improving the interfacial contact and interfacial kinetics. The success of these alloy interlayers' in SSBs has also led to their testing in liquid electrolyte based lithium metal batteries where high lithium-content Li-In ($Li_{13}In_3$) and Li-Sn ($Li_{17}Sn_4$) alloy interlayers have been shown to improve the cycling stability of metallic lithium and delay dendritic growth.[13, 14] However, the positive long-term effect of these interlayers on the anode kinetics is questionable both in SBBs and in liquid cells as the alloy interlayers may not remain located directly at the interface and the contact problem between the SE and metallic lithium electrode resurfaces.[4, 15]

Therefore, superionic conducting sulfide SEs like $Li_3PS_4$, $Li_6PS_5Cl$, or $Li_{10}GeP_2S_{12}$, which are some of the attractive SE candidates for Li-SSBs, are typically studied with bulk alloy anodes to ensure stable electrochemical performance.[16, 17] Indium metal and in particular the two phase system $xLiIn+(1-x)In$ ($x \leq 1$) with a low Li composition[18, 19] and occasionally some tin[20, 21] based alloys are the typical choice. These alloys are primarily used as the counter and/or reference electrode to facilitate electrochemical characterization of SEs, and the evaluation of cathode performance, as high reactivity of the lithium either jeopardizes or renders such assessments difficult if not impossible.[22] Surprisingly, the chemical potentials of these alloys are not compatible with the thermodynamic stability window of sulfide SEs, but stable cycling is still possible due to the formation of kinetically stable interphases.[21] The feasibility of alloys as stable anode materials has also been established for Na metal SSBs where Na-Sn alloys have been shown to kinetically stabilize the anode/electrolyte interface enabling stable operation of full cells with sulfide SEs.[23-25]

In addition to the improved chemical and electrochemical stability, Li-In and Li-Sn can potentially avoid dendritic failure and enable stable long-term cycling even at high current rates.[19, 21, 22] However, in most of these studies, the focus was rarely the investigation of the bulk alloy anodes, but rather the evaluation and optimization of the cathode or the SE performance. So far only a few theoretical[26] and experimental[9, 22] studies have looked into the fundamental behaviour of alloy anodes in SSBs, and the electrochemical performance and limitations of Li-In and Li-Sn anode compositions mostly remain unclear. Notably, in most studies, the exact alloy phase and composition are rarely probed as the alloy electrodes are prepared by simply folding and pressing In or Sn foil together with metallic lithium resulting in an uncharacterized mixture of different phases. The binary phase diagrams for Li-In and Li-Sn (see below), highlight the existence of distinct low, medium, and high lithium content phases. The often-used indium rich Li-In alloy, referred to as $Li_{0.5}In$ in literature, would not be practical beyond laboratory scale due to its poor theoretical capacity ($Li_{0.5}In$ = 113 mA h g$^{-1}$) and the relatively high and volatile price of indium (> 400 $ kg$^{-1}$).[27] Clearly, utilization of Li rich phases (i.e., a high Li/In and Li/Sn ratio) such as $Li_{13}In_3$ and $Li_{17}Sn_4$ is more desirable from a capacity and a cost standpoint.

In this context, here we unravel the feasibility of Li-In and Li-Sn alloys to serve as high-performance anode alternatives to metallic lithium in sulfide electrolyte based SSBs with an overarching motivation to provide fundamental insight into the electro-chemo-mechanics of Li-In and Li-Sn alloy systems by directly comparing their performance in symmetrical cells. For both alloy systems, a low [$Li_{0.5}In$ (xLiIn + (1-x)In, (x = 0.5)); $Li_2Sn_5$], a medium (LiIn; LiSn) and a high ($Li_{13}In_3$; $Li_{17}Sn_4$) lithium-stoichiometry phases have been synthesized and characterized with a particular focus on $Li_{13}In_3$ and $Li_{17}Sn_4$. We demonstrate the integration of the intermetallics as the bulk foil type electrode, unravel their Li-migration limitations, and probe the SE|anode interfacial contact during lithium deinsertion using $Li_6PS_5Cl$ as an example

SE. Furthermore, we show that the cell assembly and the stack pressures are important parameters and that morphological instabilities are less pronounced in alloy systems than for metallic lithium. A range of Li stripping/plating experiments coupled with computed Li migration barriers rationalize the mechanism of Li insertion/extraction in these alloys. Finally, the chemical and electrochemical stability of the alloys, particularly the high Li containing $Li_{13}In_3$ and $Li_{17}Sn_4$, is demonstrated by long term symmetric lithium cycling and impedance measurements.

## RESULTS AND DISCUSSION

### Alloy synthesis and characterization

The Li-In and Li-Sn alloys were synthesized from a stoichiometric mixture of metallic Li and In/Sn (**Figure 1a**, **Table S1**) according to the binary phase diagrams shown in **Figure 1b** and **Figure 1c**, respectively. While there are several stable intermetallic phases in both systems, for the present study a low ($Li_{0.5}In$ and $Li_2Sn_5$), a medium (LiIn and LiSn) and a high ($Li_{13}In_3$ and $Li_{17}Sn_4$) Li-stoichiometry phases were synthesized. These alloys were obtained by the solid-state synthesis in vacuum-sealed quartz ampules followed by high-energy ball milling (180 min) as schematically illustrated in **Figure 1a**. X-ray diffraction (XRD) was performed before and after the ball milling to characterize the as-obtained phases and the effect of milling. While LiIn, $Li_{13}In_3$, $Li_2Sn_5$, LiSn and $Li_{17}Sn_4$ are single phases (**Figure 1b** and **c**), the so-called $Li_{0.5}In$ is actually a solid solution of metallic In and the intermetallic phase LiIn with a total Li content of 33 atom% (xLiIn + (1-x)In; x=0.5). The powder XRD patterns of the $Li_{13}In_3$ and $Li_{17}Sn_4$ before and after ball milling are shown in **Figure 1d** and **Figure 1e,** respectively. XRD patterns for the other phases are shown in **Figure S1** and **Figure S2** of the supporting information. All patterns show an increased background at the lower angular range and broad peaks around 14, 21 and 26 °, which results from the 25 µm thick Kapton foil that was used to protect the air and moisture sensitive samples during the measurement (**Figure S1**). Except for these features, **Figure 1d** and **Figure 1e** demonstrate the phase-pure synthesis of $Li_{13}In_3$ and $Li_{17}Sn_4$, respectively, as illustrated by the good agreement with the reference patterns.[28, 29] Peak broadening can be observed in the XRD patterns for the high-Li content phases, especially in $Li_{17}Sn_4$, which most likely originates from an increase of the lattice strains (lattice imperfection) upon mechanical milling. Low Li content phases are less prone lattice strain generation under an applied mechanical force, and therefore retain the peak shape (Figure S1 and S2). In/Sn alloy electrodes were often prepared by simply pressing stoichiometric amounts of Li with In/Sn foil under high pressure.[19-22] This can lead to bulk alloy electrodes consisting of several phases, as shown on the example of Li$_{13}$In$_3$ (**Figure S3**). In contrast, our fabrication method allows the controlled and reproducible synthesis of a wide range of Li-In and Li-Sn alloys.

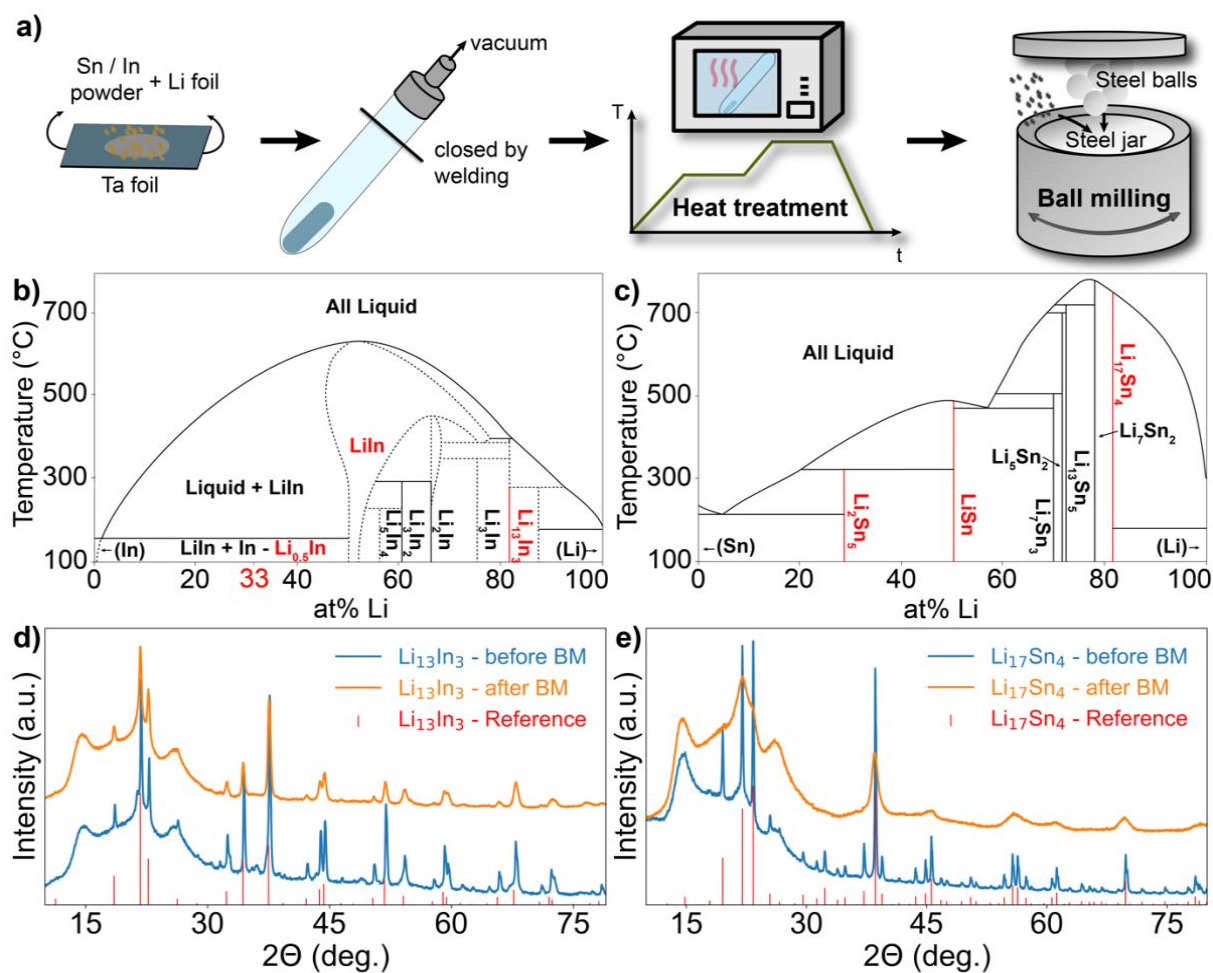

**Figure 1**. **Synthesis and characterization of the Li-alloys.** (a) Schematic representation of the synthesis protocol. (b) Binary Li-In phase diagram redrawn from Songster *et al.*[30] (c) Binary Li-Sn phase diagram redrawn from Li *et al.*[31] (d) XRD powder pattern of the Li$_{13}$In$_3$ alloy before and after ball milling (180 min) compared against the reference pattern (ICDS-no.51963)[28]. (e) XRD pattern of the Li$_{17}$Sn$_4$ alloy before and after ball milling (180 min) along with the reference pattern (ICSD-no. 240046)[29].

**Alloy processing and electrode fabrication**

Solid-state cells with a metallic lithium electrode are usually assembled by pressing a thin Li-foil directly onto the SE pellet. The applied pressure induces plastic deformation of the soft metal and ensures the desired good interfacial contact with the SE. However, the plastic deformation of hard and brittle alloys is too small to provide sufficient interfacial contact with the SE pellet. Thus, the as-synthesized alloys were first pulverized to obtain fine and uniform particles that were spread over the SE pellet. The evenly distributed powder was then pressed to form a foil-like electrode. Therefore, the high-energy ball milling step in the synthesis protocol does not only ensure reliable synthesis, but it is also necessary to grind the coarse alloy chunks (see below) obtained after the solid-state synthesis. **Figure 2a** shows the SEM micrographs of the large $Li_{13}In_3$ chunks obtained after the solid-state synthesis, and **Figure 2b** shows the fine powder obtained after ball-milling. The same pulverization was also achieved for the hard and brittle $Li_{17}Sn_4$ alloy, as evident from **Figure 2c** and **Figure 2d.** Corresponding SEM micrographs for the other In (LiIn) and Sn ($Li_2Sn_5$ and LiSn) alloys are shown in **Figure S4 and S5**, respectively. The $Li_{0.5}In$ alloy was too soft and could not be pulverized by high-energy ball milling as it got stuck on to the balls and the walls of the milling container.

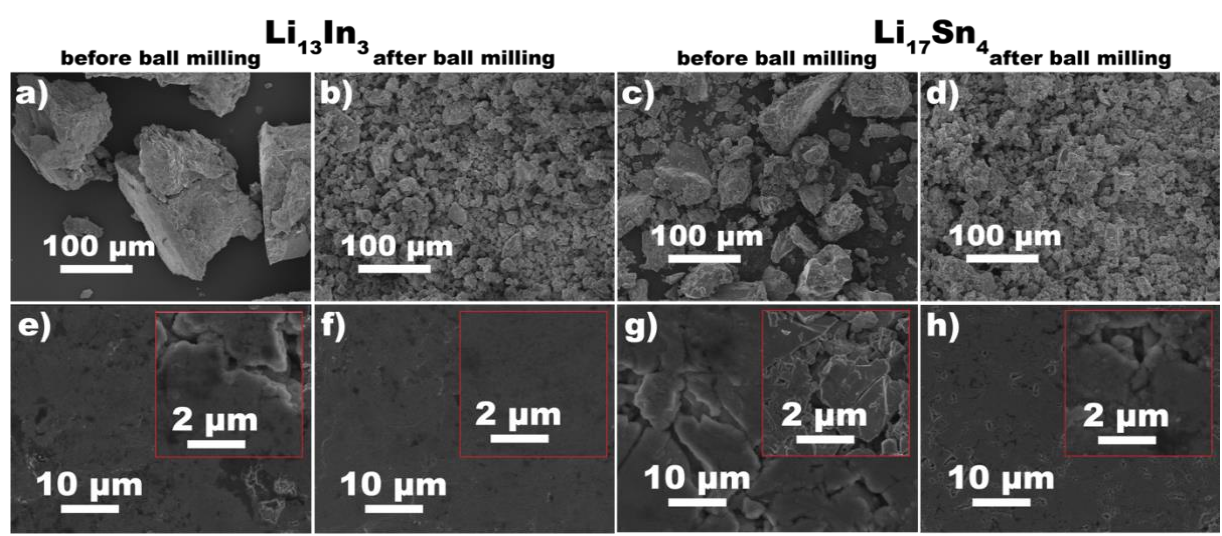

**Figure 2**. **SEM characterization of the Li-alloys.** SEM micrographs of the $Li_{13}In_3$ alloy (a) before and (b) after ball milling at 300 rpm for 3 h. SEM micrographs of the $Li_{17}Sn_4$ alloy (c) before and (d) after ball milling at 300 rpm for 3 h. SEM micrographs of $Li_{13}In_3$ pressed to a foil with (e) non-ball milled powder and (f) ball-milled powder. SEM micrographs of $Li_{17}Sn_4$ pressed to a foil with (g) non-ball milled powder and (h) ball-milled powder.

The microstructure of the alloy foils (films) obtained by pressing the coarse as-synthesized powder and the milled fine powder was investigated by SEM (**Figure 2, S4 and S5**). While the use of coarse particles leads to foils with an uneven and rough surface, the foil surface prepared from the pulverized powder appears considerably smooth. $Li_{13}In_3$ (**Figure 2e and 2f**), $Li_{17}Sn_4$ (**Figure 2g and 2h**), LiIn (Figure S4), $Li_5Sn_2$ and LiSn (Figure S5) show similar behavior with only a subtle difference arising from the difference in the malleability of the intermetallic phases. Since $Li_{0.5}In$ could not be pulverized, solid-state cells were assembled by first pressing a thin $Li_{0.5}In$ foil which was then pressed onto the SE pellet, in the same way as a Li-metal electrode (**Figure S6**). While the pressure induced deformation of the soft $Li_{0.5}In$ alloy is sufficient for optimal interfacial contact, the flat and smooth alloy electrode surface rendered by milling of the hard and brittle alloys is critical to achieve an uniform initial anode-

SE interface.

**Evolution of the anode/SE interface during Li stripping – influence of stack pressure**

To investigate the behavior of the alloy/SE interface during lithium stripping, solid-state cells (**Figure S7**) were assembled with metallic indium as the counter/reference electrode (CE), $Li_6PS_5Cl$ as the SE and metallic lithium or one of the Li-alloys ($Li_{0.5}In$, $Li_{13}In_3$ $Li_2Sn_5$ or $Li_{17}Sn_4$) as the working electrode (WE). First, the stack impedance of the assembled cell was measured. Then a constant anodic current of 200 µA was applied, and the cell potential was recorded as a function of time. The constant stripping of Li leads to its steady depletion within the WE, as indicated by an increase in the measured potential. Lastly, the stack impedance was measured again after the cell potential reached the cut-off value. For a better comparison of the stack impedance measured before and after the constant stripping experiment-coloured semicircles were used in the presented Nyquist plots. These semicircles are not based on fitted data; they are meant more as guide for the reader. In addition to metallic lithium – which was used to benchmark the behavior of the alloys – only the low ($Li_{0.5}In$ and $Li_2Sn_5$) and high ($Li_{13}In_3$ and $Li_{17}Sn_4$) lithium content alloys were analyzed as these four terminal phases perfectly summarize the behavior of these alloys in general. Alongside probing the characteristic influence of various alloys, to understand the effect of stack pressure, the stripping experiments were performed at two different stack pressures: 0 MPa and 45 MPa. To ensure equal starting conditions, all working electrodes were constructed to have a lithium capacity equivalent to ~10 mA h (**Table S2** of SI).

For an optimal CE|SE interfacial contact, the indium foil was pressed onto the SE pellet at a pressure of 150 MPa. A similar protocol was shown to enable lowly-resistive contact between metallic lithium and LLZO SE with a charge transfer resistance less than ~2 Ohm $cm^{-2}$.[32, 33] The suitability of indium as the CE was confirmed by probing the impedance of the In|$Li_6PS_5Cl$|In and $Li_{0.5}In$|$Li_6PS_5Cl$|$Li_{0.5}In$ stacks (**Figure S8**), which showed a negligibly small contribution from the interfacial resistance. Besides, the $Li_{0.x}In$ (x < 1) alloy formed upon lithiation of In (i.e., during Li stripping of the WE) provides a steady reference potential of ~0.6 V *vs*. $Li^+/Li$.[22, 34] Thus, the WE|SE|CE stack impedance, which is measured before and after the long-term stripping experiment depends only on the bulk resistance of the SE pellet ($R_{SE}$ = 38 Ohm; **Figure S**8), the WE|SE interfacial resistance ($R_{int}$) and the diffusion kinetics of the WE and CE. **Figure 3a** presents the potential profiles obtained during the stripping studies. Since all cells exhibit a different initial open-circuit voltage, we have referenced their potential to the $Li/Li^+$ scale for a better comparison of the cells' potential evolution. At a stack

pressure of 0 MPa, the metallic lithium shows a full depletion after ~7.5 h (1.5 mA h), while the various Li-alloys show a faster depletion time of ~4 min (0.01 mA h) for $Li_2Sn_5$, ~4.5 h (0.9 mA h) for $Li_{13}In_3$, ~5.5 h (1.1 mA h) for $Li_{17}Sn_4$ and ~7 h (1.4 mA h) for $Li_{0.5}In$ as highlighted in **Figure 3b**. The faster Li depletion at the SE|WE interface for the alloys decreases the total stripped capacity. Impedance measurement for the Li cell as shown in **Figure 3c** revealed a drastic rise in the interfacial impedance from the initial ~92 Ohm to ~54,029 Ohm upon stripping. These values were extracted from a qualitative fit of the Nyquist profiles with the model circuit shown in **Figure S9.** All the fitting parameters are reported in **Table S3**. It must be noted that the interfacial resistance corresponding to the Li interface rises drastically upon stripping, the resistance associated with the In interface remains almost unchanged. This observation suggests an almost complete depletion of Li at the Li|$Li_6PS_5Cl$ interface and the loss of interfacial contact as schematically illustrated in **Figure 3d**. Such a depletion mechanism is well-known and has already been elucidated in detail by Krauskopf et al.[9] **Figure 3e** shows the impedance of the $Li_{17}Sn_4$ stack before and after the striping study at 0 MPa. While the initial stack impedance of ~123 Ohm for the $Li_{17}Sn_4$ is a little higher than that for Li, the impedance after the stripping (~1738 Ohm), is over 30 times lower compared to the Li cell (**Figure 3b**). The same is true for the other alloys ($Li_{0.5}In$, $Li_{13}In_3$ or $Li_2Sn_5$) probed in the long-term stripping experiment, as shown in **Figure S10** and **Table S4**.

The striking difference in impedance outcome between the alloys and metallic Li can be explained as follows. All working electrodes were attached by briefly applying an assembly pressure of ~45 MPa to impart a good contact with the SE. But as Li is softer than the alloys, stronger plastic deformation of Li leads to much better contact with the SE and hence a lower starting interfacial resistance. The strikingly smaller interfacial impedance rise for the alloys vis-à-vis lithium after the long-term stripping can be explained by their different depletion mechanism. First of all, the lithium-alloys can be imagined as an In or Sn host matrix in which the Li is embedded. When lithium is stripped from the alloy, the host matrix still maintains sufficient contact with the SE, and therefore the impedance is only expected to increase slightly. In contrast, the Li metal forms voids, and the extensive contact loss results in a drastic increase of the interfacial impedance. Although the alloys maintain better contact with the SE, as an increasing amount of lithium is stripped, a Li concentration gradient builds up perpendicular to the interface, as schematically shown in Figure 3d, leading to a quick increase in polarization upon Li stripping. The low migration kinetics of Li in the alloys (see below) limit the replenishment of lithium from the bulk as captured by the equivalent circuit shown in

**Figure 3d**. Over time, this leads to a complete depletion of lithium in the interfacial region leaving only pure In or Sn in direct contact with the SE. Krauskopf et al. have observed and predicted a similar depletion mechanism for the $Li_{0.9+x}Mg_{0.1-x}$ alloys.[35]

To exclude any contribution from improper interfacial contact on the Li depletion mechanism, a second set of experiment was performed at a constant stack pressure of 45 MPa. It is well known that a certain stack pressure is also critical to achieve a good cathode performance and the stack-pressure of ~45 MPa is in the range applied for sulfide based SSBs in previous studies.[36, 37] For lithium, such a high external pressure can induce lithium deformation and creep, which counteracts void formation.[8, 32, 38] This allows the stripping of almost the whole available capacity of 10 mA h (**Figure 3b and S11**). For $Li_{17}Sn_4$, the 45 MPa stack pressure has two important implications. A nearly perfect interfacial contact with only ~37 Ohm stack resistance is achieved right after the cell assembly and the good interfacial contact is maintained during the whole stripping experiment as reflected by the unchanged (~38 Ohm) stack resistance after the stripping (**Figure 3f**). Other alloys display the same behavior as Li$_{17}$Sn$_4$ under the applied stack pressure as shown in **Figure S10** and **Table S4**. The stack resistance primarily corresponds to the bulk resistance of the SE pellet (R$_{SE}$ = 38 Ohm) and indicates that the WE|SE interfacial resistance is negligibly small under a stack pressure of 45 MPa. The better contact also slows the depletion and leads to a larger stripped capacity compared to the cells without any stack pressure (**Figure 3a** and **Figure 3b**). The stripped capacity increased to ~0.02 mA h for Li$_2$Sn$_5$, ~3.9 mA h for Li$_{13}$In$_3$, ~3.5 mA h for Li$_{17}$Sn$_4$ and ~3.4 mA h for Li$_{0.5}$In. These results clearly show that good interfacial contact is important to improve the performance of the alloy electrodes. In the case of lithium, the stack pressure can prevent void formation and thus enable full lithium capacity utilization. Whereas, for alloy electrodes, it is not void formation but rather the alloys' low Li migration which limits the available capacity, and this material property cannot be influenced by the stack pressure. To investigate the kinetic limitation of the alloys, their Li migration barriers were probed computationally, as presented below.

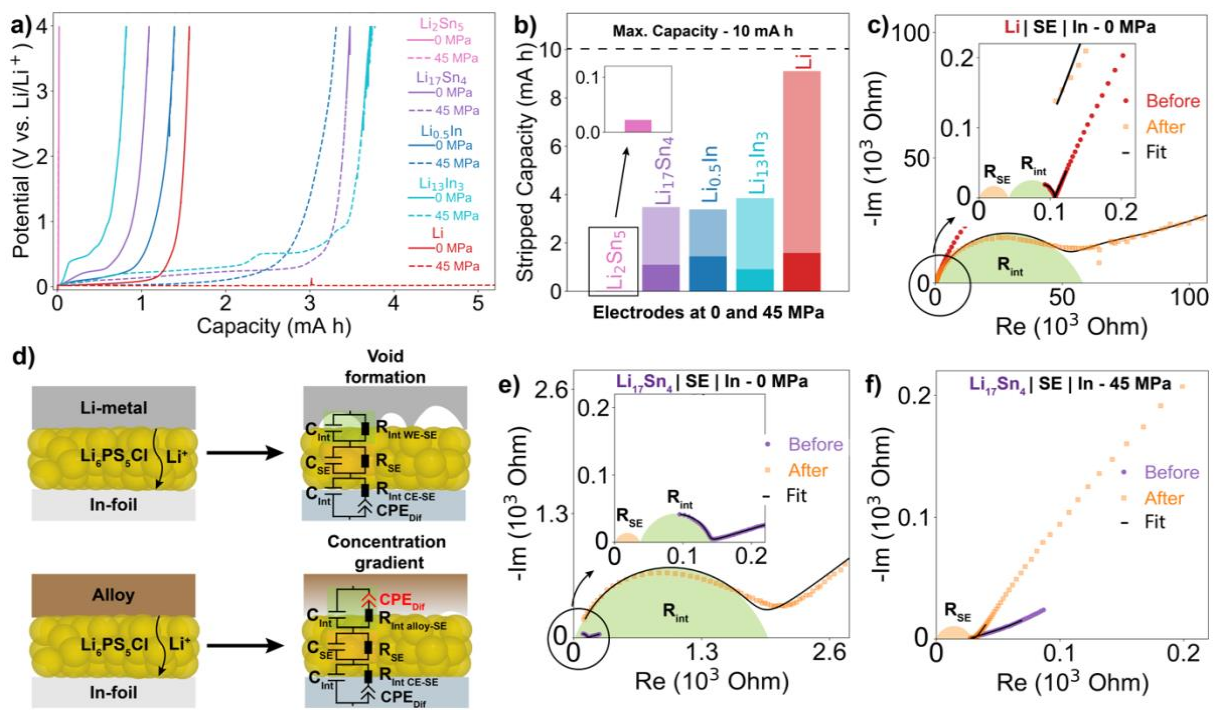

**Figure 3: Long-term stripping till depletion of the working electrode.** Long-term stripping experiment for solid-state cells assembled with an Indium counter/reference electrode and Li, $Li_{0.5}In$, $Li_{13}In_3$ and $Li_2Sn_5$, $Li_{17}Sn_4$ as working electrodes with a 750 µm thick $Li_6PS_5Cl$ pellet as the SE. (a) Potential profiles at 200 µA for the cells measured at a constant stack pressure of 0 MPa (solid line) and 45 MPa (dashed line). (b) Comparison of the stripped capacity for the different cells at 0 MPa (dark-colored bars) and 45 MPa (light-colored bars). (c) Nyquist plot of the Li cell measured at 0 MPa before (red) and after (orange) the stripping study with the fitted data shown as black line. (d) Schematic illustration and comparison of the depletion mechanism for Li and the Li-alloys as well as the impedance elements used for the data interpretation. Nyquist plot of the $Li_{17}Sn_4$ cell (e) measured at 0 MPa and (f) at 45 MPa before (purple) and after (orange) the stripping experiment with the fitted data shown as black line.

**Computational assessment of the Li migration barriers**

To evaluate the Li migration barriers of the investigated Li-In and Li-Sn phases, we first calculated the thermodynamic stability of the specific alloys and their tendency towards Li intercalation. The formation (or mixing) energy was calculated using density functional theory (DFT), as defined in **Eq. S1** in the SI. The structural models of the alloys (and the bulk metals, i.e. Li, In and Sn) were obtained from the inorganic crystal structure database (ICSD),[39] complemented by predicted structures by us as well as from the Materials Project.[40] The convex envelopes formed by the structures with the lowest formation energies, i.e. the convex hull (Eq. S.1) for the Li-In system is shown in **Figure 4a** and for the Li-Sn system in **Figure 4b.** Thermodynamically stable intermetallic phases are highlighted by the orange/green filled circles, whereas violet crosses represent metastable or unstable phases.

In **Figure 4a**, a number of stable compounds are observed, which are: LiIn$_3$ (tetragonal), Li$_5$In$_4$ (trigonal), Li$_3$In$_2$ (rhombohedral), Li$_2$In (orthorhombic), and Li$_{13}$In$_3$ (cubic), from the Li-In convex hull (see **Table S5**). The convex hull hits its global minima at –293.4 meV/atom for Li$_3$In$_2$. Both the cubic LiIn and Li$_3$In appear slightly metastable from our predictions (~3.0 and ~2.0 meV/atom above the convex hull, respectively), but these phases are present in the experimental phase diagram (**Figure 1b**) and should be considered stable even at low temperatures.[41] In contrast, LiIn$_2$ (i.e., Li$_{0.5}$In), a popular alloy anode electrode that is also investigated in this report remains about ~138 meV/atom above the convex hull, in agreement with predictions from the MaterialsProject (~140 meV/atom). A better representation of a slightly lithiated In electrode is given by LiIn$_{35}$ (which was obtained by searching the Li-In tie line and appears only ~24 meV/atom above the convex hull (and below the thermal energy at ~30 °C of 25 meV/atom). Thus, it is relevant to investigate the Li transport in LiIn$_{35}$ as a proxy for lowly lithiated In electrodes like Li$_{0.5}$In alloy, primarily used experimentally.

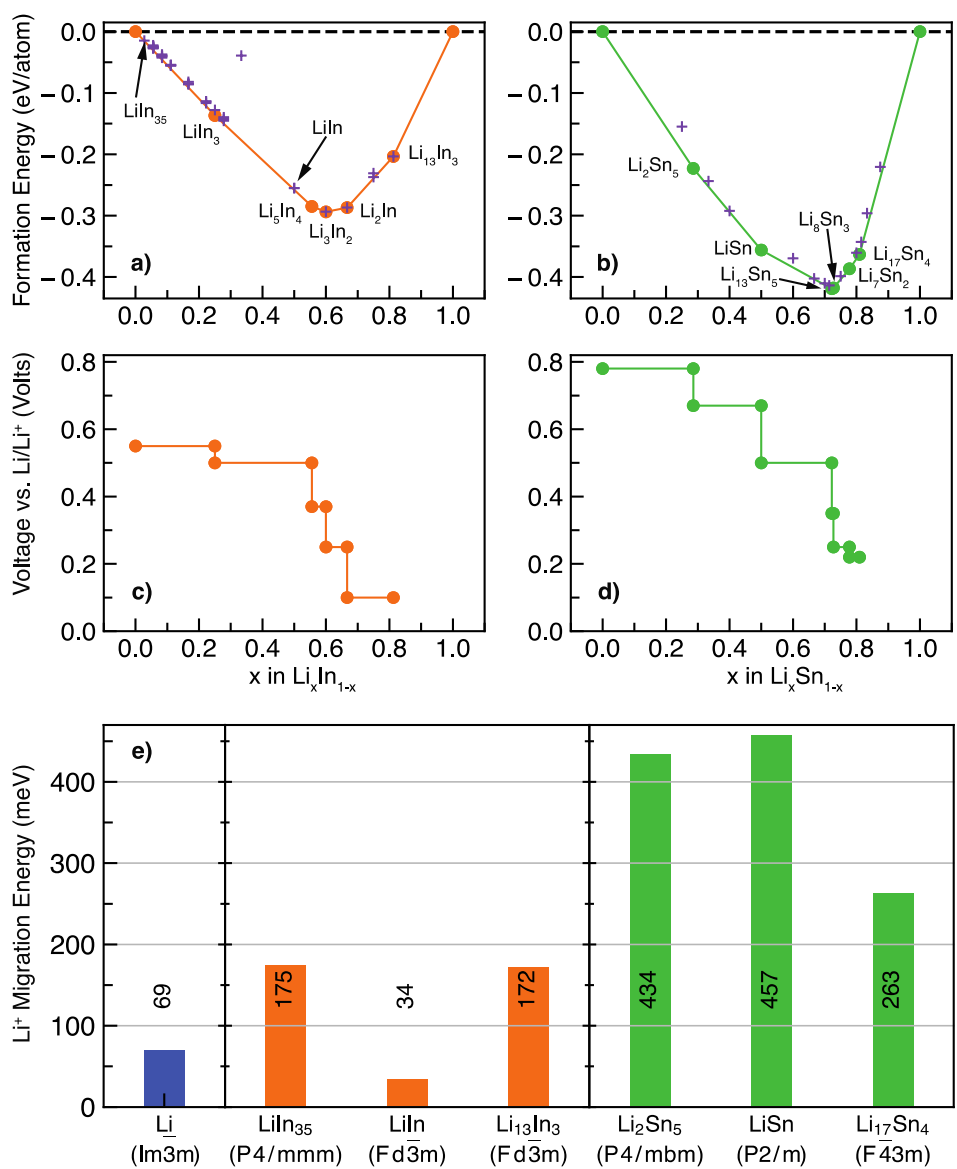

**Figure 4**. **Computational results for the Li-In and Li-Sn alloys.** Formation (mixing) energy diagram for Li intercalation in (a) In and (b) Sn. Stable (ground state) structures are shown by orange/green circles and metastable/unstable structures by purple crosses, respectively. Computed voltage curves for Li intercalation in (c) In and (d) Sn. (e) Computed Li$^+$ migration energies in Li-In and Li-Sn alloys. Only the lowest migration energies associated to percolating Li$^+$ paths in each alloy are shown. The computed migration energy paths are reported in **Figure S12** of the Supporting Information. The metastable compound LiIn$_{35}$ in (a) is used as a proxy to study regimes of low lithiation.

We extended the analysis to the Li-Sn system (**Figure 4b**), which reproduced existing experimental [42, 43] and computational [44-46] reports. Here, the convex hull is formed by $Li_2Sn_5$ (tetragonal), LiSn (monoclinic), $Li_{13}Sn_5$ (trigonal), $Li_8Sn_3$ (rhombohedral), $Li_7Sn_2$ (orthorhombic) and $Li_{17}Sn_4$ (cubic), with a deep minimum at –419.0 meV/atom associated with $Li_{13}Sn_5$ (**Table S5**). An extensive assessment of Li intercalation in In and Sn alloys is not in the scope of this work, which has been thoroughly discussed in some excellent reports earlier.[22, 47, 42, 43] Yet, for completeness, we briefly present the non-topotactic voltage curve for the Li-In system in **Figure 4c** and the Li-Sn system in **Figure 4d,** derived from the convex hulls of **Figure 4a** and **Figure 4b**, respectively. Indium allows the insertion of ~4.34 Li and an average voltage of ~0.55 Volts vs. $Li/Li^+$. Each minimum in the convex hull (**Figure 4a**) represents a step in the voltage curve, separated by voltage plateaus – an indication of two phase-regions in the phase diagram. Likewise, a non-topotactic voltage curve for the Li-Sn alloy reveals the insertion of ~4.25 Li in Sn and an average voltage of ~0.60 Volts vs. $Li/Li^+$.

The experimentally studied (this work) Li-In and Li-Sn alloys which are identified by the computational search were studied for the Li migration in the regime of dilute vacancy, i.e., one defect per supercell. Using a combination of DFT and nudged elastic band, the Li migration barriers presented in **Figure 4e** were computed for a subset of alloys (in addition to Li-bulk), which include, $LiIn_{35}$, LiIn, $Li_{13}In_3$, $Li_2Sn_5$, LiSn, and $Li_{17}Sn_4$. The lowest migration energy paths enabling Li percolation are shown in **Figure S12**. Specific alloys provide other alternative Li migration pathways that lead to higher migration energies shown in **Figures S13-S15**.

As expected, the migration of Li in Li metal (bulk) shows the lowest migration energy (~69 meV, Figure 4e) among all the alloys considered. An exception to this trend is LiIn (~34 meV), where the facile Li migration occurs between two face-sharing distorted $LiIn_4$ tetrahedra (see **Figure S13f**). Although, experimentally LiIn is reported as a stable compound at room temperature (**Figure 1b**), LiIn appears slightly metastable (and located ~3 meV/atom above the stability line) in the computed phase diagram at 0K. Since the end point structures in our NEB calculations sites are indeed metastable, the computed migration barriers at 0K are artificially lowered. Considering the thermodynamic instability of our initial and final structure models (and that the model of LiIn contains 127 atoms), a decrease of stability of the end-point structures of at least ~381 meV is expected. Therefore, if the metastability of LiIn is accounted, the $Li^+$ migration barrier will be ~ 415 meV (381 + 34 meV), which is nearly 3 times higher than the barriers computed in the case $Li_{13}In_3$ (~172 meV, **Figure 4**). More computational and experimental investigations are required to fully assess the migration characteristics of Li$^+$ in LiIn.

The migration energy barriers in Li-Sn systems are comparatively higher than the corresponding values for Li-In systems, which could be linked to the difference in atomic radius between In (~1.55 Å) and Sn (~1.45 Å), leading to a wider bottleneck for Li migration in Li-In alloys. All the other Li-In and Li-Sn alloys show energy barriers below 500 meV, which appear reasonable in the context of typical migration energies measured and computed in other anode materials. For example, the Li migration barrier in graphite has been estimated to range between 208-400 meV.[48] To give more context, it has been estimated that a barrier of 525 meV is equivalent to a diffusivity of $10^{-12}$ cm$^2$ s$^{-1}$ for a 1 mm particle size of active material at a discharge rate of 2C.[49, 50]

Experimentally, LiIn shows lower overpotentials for Li cycling (plating/stripping) than Li$_{13}$In$_3$ (albeit at lower currents, see below), which agree well with the computed Li-migration barriers of ~34 meV and ~172 meV, respectively. Similarly, the experimentally observed Li cycling overpotentials for (see below) the Sn alloys – Li$_2$Sn$_5$ > LiSn >> Li$_{17}$Sn$_4$ – follow the simulated migration barriers trend (~434 ≈ ~457 > ~263 meV, respectively). Each Li atom is coordinated by 8 and 10 Sn atoms in Li$_2$Sn$_5$ and LiSn, respectively. In contrast, in Li$_{17}$Sn$_4$, Li coordination is 6 or 8, which seems to suggest that the lower Li coordination number facilitates Li migration in Li$_{17}$Sn$_4$ alloys. While the lowest barrier for Li migration in Li$_2$Sn$_5$ is ~434 meV, another migration mechanism (**Figure S14**) was computed, entailing a much larger barrier of ~1208 meV. Likewise, other possible migration mechanisms in LiSn correspond to barriers of ~500 and 900 meV.

Recently, Qu et al.[46] have studied the Li transport in Li-In and Li-Sn alloys and proposed two distinct mechanisms implying the migration of Li interstitials or vacancies, with the latter discussed in our study. While the absolute values of migration energies computed in their study agree well with our simulations, they pointed out a change in mechanism from an interstitially mediated to a vacancy-mediated one with increasing Li concentration in the alloys. Nevertheless, the migration barrier is also a function of the relative stability of the initial and arrival Li sites in these alloys. Therefore, any interstitial site with appreciable thermodynamic instability may lead to decreased migration barriers. Finally, we speculate that an interstitial-type migration mechanism should be preferred for low-Li content alloys, in contrast to the hypothesis put forward by Qu et al.[46]

In light of our findings, we believe that Li percolation in the alloy upon electrochemical lithiation must follow a combination of migration paths depending on the local coordination environment experienced by Li. We have shown that specific migration mechanisms entail non-negligible migration barriers, which correlate (except in the case of Li migration in LiIn) with an increase in the experimentally observed overpotential. Furthermore, the local coordination imposed by the host alloy on Li (see **Table S6**) influences the Li migration. In both type of alloys considered, i.e., Li-In and Li-Sn, migration paths involving Li departing from sites with large coordination numbers (8-10), set by the alloying element, consistently provide large migration energies, in comparison to lower coordination numbers (6-4). This notion is in striking contrast to the typical design rules of Li migration in oxide and sulphide host materials.[50] In particular, one of the design principles in these host materials suggests that when Li ions are found in undesired anion coordination environments –everything different from tetrahedra or octahedra– Li migration barriers decrease and ion transport is facilitated.[50] This phenomenological observation is in contrast to the results on our simulation in the Li-In and Li-Sn alloys, an aspect that warrants more future investigation.

**Li cycling behavior in symmetric cells**

We have seen above that during electrochemical Li stripping, the Li-alloys show a different depletion mechanism compared to metallic lithium, but the same mechanism is also expected to influence the plating (lithium incorporation) behavior of the alloys. To understand the behavior under reversible Li depletion/incorporation regime, the alloys were studied by galvanostatic stripping/plating in a symmetrical A|Li$_6$PS$_5$Cl|A (A = Li$_{0.5}$In, LiIn, Li$_{13}$In$_3$, Li$_2$Sn$_5$, LiSn or Li$_{17}$Sn$_4$) configuration by applying currents ranging from 50 – 1000 µA cm$^{-2}$ (capacity: 0.05 – 1 mA h cm$^{-2}$) (**Figure 5**). The cells were assembled by pressing the alloy on both sides of the SE pellet with an assembly pressure of 150 MPa. A stack resistance of around ~40 Ohm (where R$_{SE}$ = 38 Ohm, **Figure S16** and **Table S7**) for all assembled cells verified the formation of nearly perfect initial contact, which was maintained during the repeated Li cycling by applying a constant stack pressure of 45 MPa. However, a much smaller assembly and stack pressure had to be applied for the reference Li cell, as its behavior is dominated by creep and extrusion of Li through the pores of the SE (**Figure S17**), as we reported earlier.[11] The galvanostatic Li cycling for the alloys leads to only a small rise in impedance (**Figure S16** and **Table S7**), which highlights the excellent chemical and electrochemical stability of the alloys in contact with Li$_6$PS$_5$Cl. This finding also suggests that the interfacial phenomena cannot account for the variation in polarization observed during the Li cycling of the alloys (see below). A better comparison of the voltage polarization of the alloys can be achieved by dividing the measured overpotential into two parts: (i) the minimum expected overpotential ($U_{min}$) corresponding to the resistance of the SE ($R_{SE}$) and the interface ($R_{int}$), i.e., $U_{min} = I \cdot R_{min}$ with $R_{min} = R_{SE} + R_{int} = $ ~40 Ohm (**Table S7**) and (ii) the kinetic overpotential arising from the Li migration barriers of the alloys. In **Figure 5a-f**, the minimum expected potential ($U_{min}$) is denoted by the red dashed lines.

**Figure 5a** shows the potential profile for the symmetric $Li_{0.5}In$ cell during the various constant current Li cycling. As the electrode potential is independent of the Li composition variation in the cycled capacity regime (from 0.1 to 1 mA h cm$^{-2}$) – shown by the calculated voltage curve for the Li-In system in **Figure 4c** - a plateau-like potential profile is observed here. The voltage polarization corresponds to the minimal expected overpotential (corresponding to $R_{SE}$; see above) and increases linearly with the applied current, suggesting that the lithium transport kinetics is not a limitation, as expected for a biphasic (de)lithiation mechanism. The LiIn cell, on the other hand, displays a sloped potential profile with polarization much above the minimum expected overpotentials (**Figure 5b**). This can be explained by the drift of the electrode potential with increasing Li composition, as shown by the calculated voltage curve for the Li-In system in Figure 4c. At 100 µA cm$^{-2}$ (0.1 mA h cm$^{-2}$) the overpotential of the LiIn cell is ~38 mV (at the top of the slope), which is over seven times higher than the overpotential observed for the $Li_{0.5}In$ cell (~5 mV) at the same current. When the current is increased to 1000 µA cm$^{-2}$ (1 mA cm$^{-2}$), the LiIn cell displays a large jump in the potential profile. This is not surprising as the large compositional change (large x in $Li_{1+x}In$) accompanying a substantially large capacity of 1 mA h cm$^{-2}$ (compared to 0.1 mA h cm$^{-2}$ at 100 uA cm$^{-2}$) results in a large drift in the electrochemical potential of the alloy anode undergoing electrochemical (de)lithiation. The computed voltage for the Li-In system as presented in **Figure 4c** shows that the LiIn phase is located near a voltage step between $LiIn_3$ and $Li_5In_4$. Therefore, we believe that the measured potential profile at low currents (100 µA cm$^{-2}$) and at low stripped/plated capacities are better suited to compare the Li-transport kinetics of the various alloys. Furthermore, the small migration barrier for LiIn is due to its metastability.

Compared to the $Li_{0.5}In$ and LiIn phases, the high-lithium containing $Li_{13}In_3$ shows the highest overpotential of ~95 mV at 100 µA cm$^{-2}$ (**Figure 5c**). However, when the current is increased to 1 mA cm$^{-2}$ the potential only increases to ~170 mV, which is three times lower than the observed ~520 mV for LiIn at the same current. The computed voltage diagram for the Li-In system in **Figure 4c** demonstrates that the potential for the high lithium-containing phases with an atomic lithium fraction larger than 0.7 does not change significantly and is around 100 mV vs. Li/Li$^+$. Therefore, even at large stripped/plated capacities of 1 mA h cm$^{-2}$, the electrode potential does not drift much, resulting in a rather flat voltage profile at 1 mA cm$^{-2}$ without cell failure or full Li-depletion of the electrode interface region, which would result in an exponential increase in the overpotential. Clearly, the migration barrier increases with increasing Li content and all three alloys (LiIn$_{35}$, LiIn, Li$_{13}$In$_3$) exhibit moderate to very good Li-transport kinetics, further supported by our computed Li migration barriers (see above).

The symmetric Li cycling study of Li-Sn alloy electrodes unravels a different story. **Figure 5d** shows the potential profile for the Li cycling of a symmetric Li$_2$Sn$_5$ cell. Unlike the Li-In alloys, the potential increases exponentially and reaches the set potential limits of ± 2 V already at 100 µA cm$^{-2}$. Even at a small 50 µA cm$^{-2}$ the overpotential rises/drops to ± 0.8 V. A similar behaviour is observed for the symmetrical LiSn cell as shown in **Figure 5e**. The potential touches the cut off limit at a current of 250 µA cm$^{-2}$. The computed voltage-composition plot in **Figure 4d** indicates that the extraction/insertion of Li from/into these two phases (i.e., Li$_2$Sn$_5$ and LiSn) and consequent compositional variation will result in a constant but small change in potential (± 0.2 V). Thus, the electrode potential drift alone cannot explain the extremely large polarization at small currents, leading to the assumption that low lithium content Sn-phases possess extremely poor Li-transport kinetics, which is supported by our computed high Li migration values of 434 meV and 457 meV for Li$_2$Sn$_5$ and LiSn, respectively. In comparison, a faster Li migration in the Li$_{17}$Sn$_4$ (263 meV) allows its reversible Li cycling even at 1000 µA cm$^{-2}$ (**Figure 5f**). Yet, compared to Li$_{13}$In$_3$, Li$_{17}$Sn$_4$ exhibits a rather large and sloping polarization. The significantly high Li migration barrier of Li$_{17}$Sn$_4$ compared to Li$_{13}$In$_3$ (172 meV) can explain the higher Li cycling overpotential of the former, but the origin of the sloping potential is not clear from the computed potential-composition profile as the composition change around Li$_{17}$Sn$_4$ is expected to lead to only a small drift in electrode potential as for Li$_{13}$In$_3$.

Overall, the galvanostatic Li cycling data of the symmetrical cells lead to the conclusion that the Li-In alloys exhibit better Li transport kinetics than the Li-Sn alloys. A comparison of the potential profiles of the alloy phases at 100 µA cm$^{-2}$ suggests that the Li transport kinetics in the Li-In system decreases with increasing Li content (Li$_{0.5}$In > LiIn > Li$_{13}$In$_3$). Interestingly, the opposite behavior is observed for the Li-Sn system, i.e., the Li transport kinetics increases with increasing Li content (Li$_2$Sn$_5$ < LiSn < Li$_{17}$Sn$_4$). However, since not only the Li transport kinetics but also the drifting electrode potentials with changing composition directly affect the observed polarizations, it is difficult to deduce the alloys' exact Li transport kinetics from the measured potential profiles. In this regard, the computed Li migration barriers and potential-composition data are critical to rationalize the experimental observations. It is important to note that the Li cycling potential in the symmetric cycling data typically includes equal contributions from both electrodes, and therefore the polarization contribution by any of the alloy anodes to a full cell would be only half of that observed here: 85 mV for Li$_{13}$In$_3$ and 210 mV for Li$_{17}$Sn$_4$ for 1 mA - mA h cm$^{-2}$ cycling. Such moderate overpotentials in combination with high lithium utilization would be attractive if long term stable cycling is delivered, and that is what we investigate below.

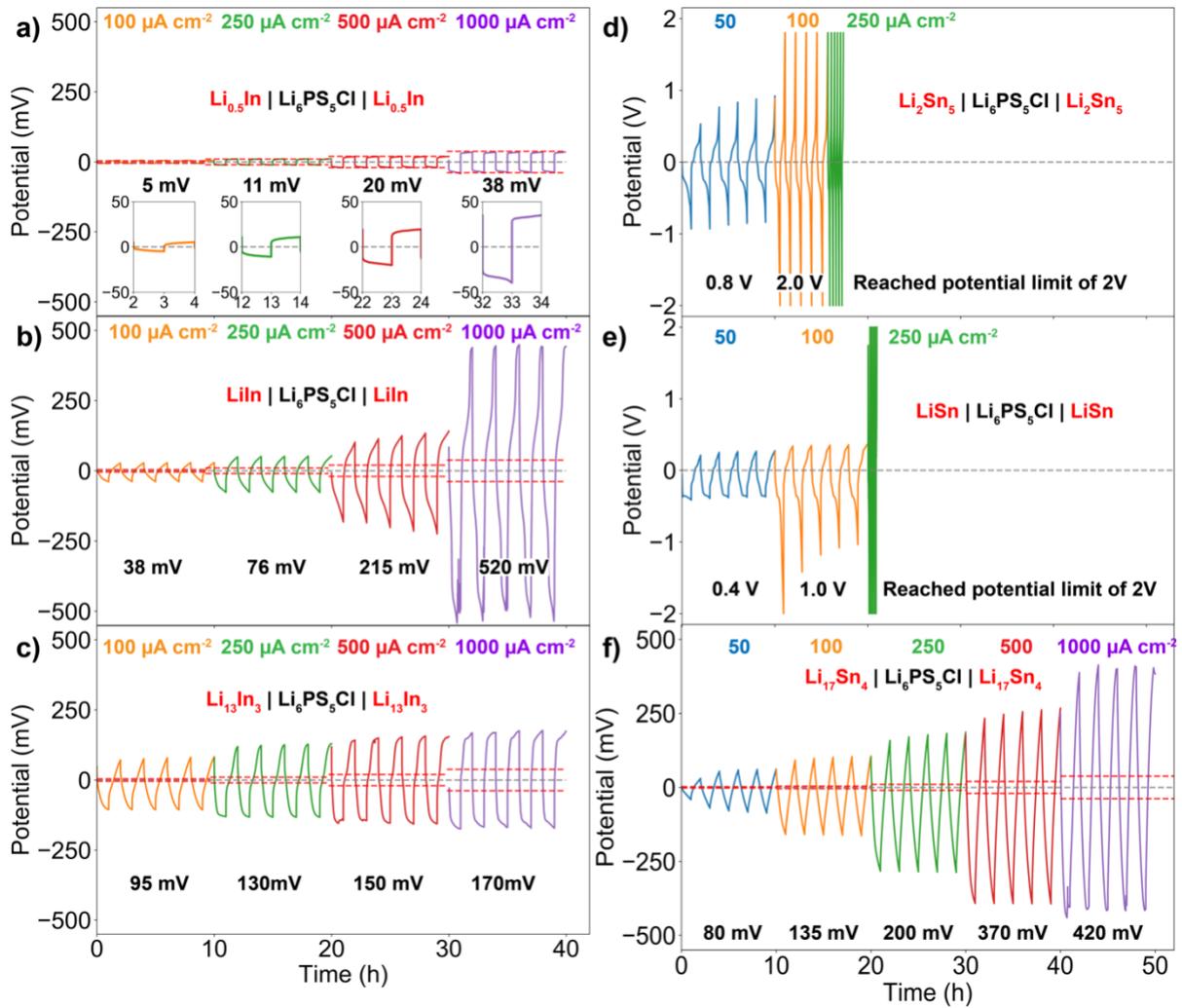

**Figure 5**. **Stripping and plating performance of the different alloys.** Li stripping and plating performance of symmetrical A|Li$_6$PS$_5$Cl|A (A = Li$_{0.5}$In, LiIn, Li$_{13}$In$_3$, Li$_2$Sn$_5$, LiSn or Li$_{17}$Sn$_4$) cells at a stack pressure of 45 MP with a potential cut-off of ±2 V for various currents: 50, 100, 250, 500 to 1000 µA cm$^{-2}$ and 1 h plating/stripping duration. Recorded potential profiles for the galvanostatic cycling of the symmetrical cell with (a) Li$_{0.5}$In, (b) LiIn, (c) Li$_{13}$In$_3$, (d) Li$_2$Sn$_5$, (e) LiSn and (f) Li$_{17}$Sn$_4$. The red dashed line in the polarization profiles correspond to the overpotential (IR$_{min}$, see text) expected based on the resistance of the electrolyte pellet (R$_{SE}$) and the interfacial resistance (R$_{int}$).

**Long-term Li cycling performance**

Long-term Li cycling performance of the Li-In and Li-Sn alloys was investigated in a symmetric cell configuration with $Li_6PS_5Cl$ as the SE by applying a constant current of 1 mA $cm^{-2}$ for a lithiation/delithiation duration of 1 h (equivalent to 1 mA h $cm^{-2}$ capacity). While the set current and cycled capacity (per $cm^{-2}$) values are not optimal for practical applications, these are considerably high and reveal a reliable performance of the alloy anodes.

The results for high Li content phases —i.e., $Li_{13}In_3$ and $Li_{17}Sn_4$, which are more practical solid-state Li-anode candidates from cost and capacity considerations - are discussed here in detail. The $Li_{0.5}In$ and LiIn alloys' performance are presented in **Figure S18**. As evident from **Figure 6a**, the symmetric $Li_{13}In_3$ cell displays a steady overpotential (~180 mV) for the 1000 h or 500 cycles probed here. A better overview of the entire cycling period is shown in **Figure S19**. Noticeably, individual potential profiles, shown in the inset of **Figure 6a**, display almost identical plateau-like features, indicating the long-term chemical and electrochemical stability of $Li_{13}In_3$ with $Li_6PS_5Cl$ as the SE. The stack impedance measured before and after the long-term cycling experiment (**Figure 6b**) reveals only a minor increase in the stack impedance from ~38 Ohm to ~54 Ohm, which further confirms the compatibility of the $Li_{13}In_3$ anode with the SE. The small impedance rise most likely stems from a slight loss in contact after 1000 h. A similar increase in stack impedance was also observed for the $Li_{0.5}In$ and LiIn cells, which also show stable long-term Li cycling behavior (**Figure S18**).

Like $Li_{13}In_3$, $Li_{17}Sn_4$ too displays a long-term Li cyclability (**Figure 6c**), albeit with a higher overpotential which also increases from ~440 mV to ~650 mV in the course of the cycling. However, the overpotential rises only within the first ~300 h and then stabilizes, as shown in **Figure S19**, most likely due to the formation of a stable passivation layer at the $Li_{17}Sn_4$-SE interface. The stack impedance measured before and after the cycling study shows an increase from ~38 Ohm to ~65 Ohm (**Figure 6d**), which alone cannot explain ~200 mV increase in overpotential. The Li-transport kinetics likely changes over time with the repeated insertion and expulsion of the Li, which leads to the observed change in the potential profile (inset **Figure 6c**). At the beginning (left inset; 25 h), the potential profile appears curved and rises constantly during each half-cycle. However, after more than 500 hours, the potential rises rather sharply at the onset of the stripping/plating before slowly flattening out. The rapid change in potential at the beginning of the polarization cycle indicates a deteriorated

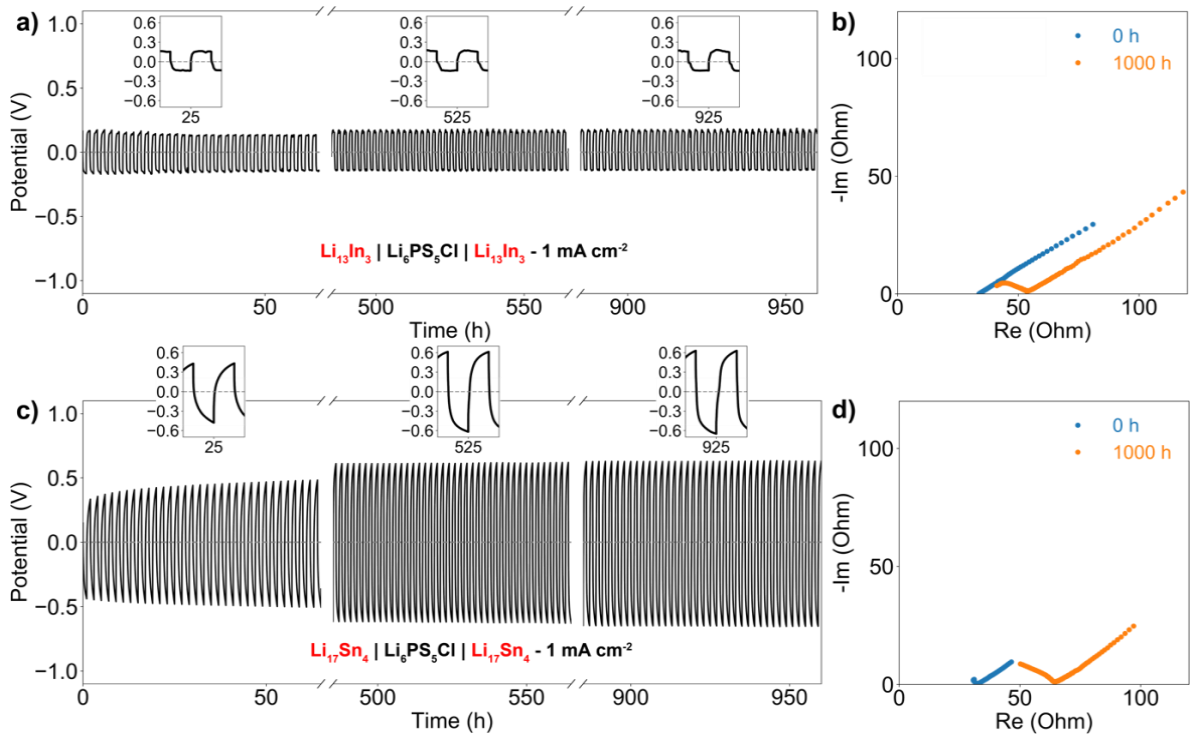

**Figure 6. Long-term Li cycling performance.** Repeated Li cycling of symmetrical cells at a stack pressure of 45 MPa, a constant current of 1 mA cm$^{-2}$ and a capacity of 1 mA h cm$^{-2}$. (a) The potential profiles for the Li$_{13}$In$_3$|Li$_6$PS$_5$Cl|Li$_{13}$In$_3$ cell and (b) the corresponding Nyquist impedance profiles recorded before and after the stripping experiment. (c) The polarization profile for the Li$_{17}$Sn$_4$|Li$_6$PS$_5$Cl|Li$_{17}$Sn$_4$ cell and (d) the corresponding Nyquist impedance plots recorded before and after the cycling study.

Li-transport kinetics induced by repeated (de)lithiation. One possible explanation could be the formation of small local areas of low Li-containing phases at the interface, which exhibit poor Li-transport kinetics (see above). Nevertheless, Li$_{17}$Sn$_4$ exhibits good chemical and electrochemical stability with Li$_6$PS$_5$Cl to allow repeated Li cycling at 1 mA cm$^{-2}$ and 1 mA h cm$^{-2}$ for the 1000 h probed here.

**CONCLUSION**

Morphological instabilities and loss of contact at the anode - solid-electrolyte interface are critical challenges in the practical implementation of lithium metal anode in solid-state batteries. The application of sufficient stack pressure can avert the problem. Yet, creep and extrusion of soft Li through the micropores of the electrolyte membrane, even under a moderate stack pressure, can induce dendritic short at very low current densities. Bulk Li-In and Li-Sn alloy electrodes, prepared and fabricated in a controlled manner, can mitigate both the morphological instability and extrusion related problems and ensure an excellent interfacial contact during the solid-state battery operation. Furthermore, unlike metallic Li, the intermetallic phases - including those with high Li content Li$_{13}$In$_3$ and Li$_{17}$Sn$_4$ - possess impressive (electro)chemical stability in contact with the sulfide solid electrolyte $Li_6PS_5Cl$. The chemical stability and stable interfacial contact aided by an optimal stack pressure allow stable long-term Li cycling at relatively high currents (1 mA cm$^{-2}$) and capacities (1 mAh cm$^{-2}$). However, compared to Li metal, the alloys suffer from relatively slow Li migration, which not only leads to a higher Li cycling overpotential but may also result in a lower Li utilization. The complementary experimental and theoretical results further reveal the increase of the Li migration barrier with increasing Li content in the Li-In phases and decreasing Li content in the Li-Sn phases, and thus provide fundamental insight into the electrochemical behavior of the bulk alloy electrodes in SSBs. High lithium containing bulk alloy anodes are clearly attractive from a cost and Li capacity point of view, but full capacity utilization might not be possible at room temperature owing to the kinetic limitation and likely significant volume changes. A feasible solution is the use of nanoalloys and composite anodes, but SE incorporation would need to be optimized so as not to sacrifice the gravimetric and volumetric capacities. Overall, we believe the presented understanding will provide a guideline in the development of alternative high-performance alloy anodes for SSBs and may help in developing strategies to overcome the electro-chemo-mechanical issues at the anode-SE interface, which are arguably one of the biggest challenges in the integration of anode materials in SSBs.

**EXPERIMENTAL SECTION**

**Synthesis of $Li_6PS_5Cl$.** $Li_6PS_5Cl$ was prepared by the well-established ball milling and heat-treatment mediated approach (see supporting information).

**Synthesis of the lithium alloys.** The various alloys [$Li_{0.5}$In (In + LiIn), LiIn, $Li_{13}In_3$, $Li_2Sn_5$, LiSn and $Li_{17}Sn_4$] were prepared by a solid-state reaction of the pure metals followed by high energy ball milling. For the synthesis, 0.5 g batches of stoichiometric amounts Li and In or Li and Sn (see Table S1) were pressed together between two stainless-steel plates at 150 bar. The deformable mass was scratched from the steel plates and wrapped into tantalum foil. The wrapped package was sealed in a carbon coated quartz ampoule under 10$^{-5}$ bar pressure. The sealed samples were processed by multistep heat treatment. In the first step, the samples were heated at 200 °C for 12 h (heating rate 20 °C h$^{-1}$) than heated at 300 °C for 12 h (heating rate 20 °C h$^{-1}$) and finally heated at 400 °C for 12 h (heating rate 20 °C h$^{-1}$) which results in a 48 h multistep heat treatment. The sealed quartz tubes were opened in an argon-filled glovebox and labelled as non-ball-milled samples. Half of each sample was then ball-milled in stainless-steel jars with 10 steel balls (10 mm in diameter) at 300 rpm. The alloys were milled for 42 min (5 min of milling, 2 min pause), which corresponds to a total milling time of 30 min. After 30 minutes of milling, the jars were opened in the glovebox, and the powder was scratched off the wall. This procedure was repeated till a total milling time of 180 min was reached.

**Computational methodology.** Details on the computational methodology is described in the supporting information.

**Electrochemical characterization.** The electrolyte pellets were prepared with a home-made Nylon die mold (d = 12 mm) and harden steel rods by pressing 150 mg (if not further specified) of the electrolyte powder at a fabrication pressure of 510 MPa for 1 min with a hydraulic press. The ionic conductivity of the electrolyte pellets was evaluated by impedance spectroscopy (EIS) using a two-probe ac impedance spectroscopy analyzer (SP200, Biologic) in the frequency range from 7 MHz to 100 mHz with a potentiostatic signal perturbation of 10 mV. For the EIS measurement, the cells were clamped inside our home-made pressure vice and closed with a torque-wrench to control the stack-pressure during the measurement. For the assembly of symmetric A|$Li_6PS_5Cl$|A cells (A = alloy or metallic lithium) SE pellets were prepared as explained above, and the full stack was assembled according to the specific alloy used as electrodes in the cells. For the symmetric Li|$Li_6PS_5Cl$|Li cells, lithium metal was first cleaned with a razor blade and then pressed onto the steel rod (current collector) with 150 MPa in order to get a thin metal disk with a flat surface, and the protruded metal was cut off. Then the rods were pushed in the Nylon die, and the solid-state cells were closed without applying any pressure (assembly pressure = 0 MPa). Symmetric $Li_{0.5}In$|$Li_6PS_5Cl$|$Li_{0.5}In$, or In|$Li_6PS_5Cl$|In cells were prepared by pressing $Li_{0.5}In$ or In alloy onto the steel rod with 150 MPa in order to get a thin metal disk and protruded metal was cut off. Then the rods were pushed in the Nylon die, and the solid-state cells were closed by applying an assembly pressure of 150 MPa. For the assembly of A|$Li_6PS_5Cl$|A cells (A = LiIn, $Li_{13}In_3$, $Li_2Sn_5$, LiSn and $Li_{17}Sn_4$), about 10 mA h (Table S2) equivalent of the respective Li-alloy was spread evenly on both sides of the SE pellet and the solid-state cells were closed by applying an assembly pressure of 150 MPa.

The stack impedance of the A|$Li_6PS_5Cl$|A cells (A = alloy or metallic lithium) was evaluated by impedance spectroscopy (EIS) using a two-probe ac impedance spectroscopy analyzer (SP200, Biologic) in the frequency range from 7 MHz to 100 mHz with a potentiostatic signal perturbation of 10 mV). For the EIS measurement, the cells were clamped inside our home-made pressure vice and closed with a pressure of 45 MPa using a torque-wrench to control the stack-pressure during the measurement. Metal stripping and plating (VMP3, Biologic) was

performed in a symmetrical cell set-up with two Li-metal electrodes. The stack pressure was then set to 45 MPa (alloy) or 5 MPa (metallic Li) during the measurement. The Li stripping and plating performance was performed in 1 h step at a fixed current of ± 50, 100, 250, 500 or 1000 µA cm$^{-2}$ at room temperature.

## ASSOCIATED CONTENT

### Supporting Information

Supplemental information contains additional experimental details, the diffraction pattern of the Li-In and Li-Sn alloys, SEM micrographs of the alloys and the fabricated electrodes, the resistance of the SE pellets, the stack resistance of A|Li$_6$PS$_5$Cl|A (A= Li, In, Li$_{0.5}$In, LiIn, Li$_{13}$In$_3$ Li$_2$Sn$_5$ or Li$_{17}$Sn$_4$) cells, the stripping and plating performance of metallic lithium and the long-term stripping and plating performance of cells assembled with Li$_{0.5}$In and LiIn. The computational details, a list of the compound studied with modeling, their provenance, the definition of the formation energy, and the additional Li migration barriers in selected alloys are also available in the Supporting information.

### Acknowledgements

D. K. and C. H. acknowledge the Swiss National Science Foundation (SNSF) for the financial support for this work through their Ambizione grant. D. K. also acknowledges the UNSW for the support through the start-up grant.  P. C. and B. S. acknowledge funding from the National Research Foundation under his NRF Fellowship NRFF12-2020-0012. The computational work was performed on resources of the National Supercomputing Centre, Singapore (https://www.nscc.sg).

### Author Contributions

D. K. and C. H. conceived the project and designed the experiments. C. H. performed all the experimental work with help from D. K. (David) and analyzed the data. B. S. and P. C. carried out the computation investigation and the data analysis. D. K. and C. H. wrote the manuscript with help from all authors.

### Conflict of Interests

There is no conflict of interests to declare.

**TOC Graphic (For Table of Contents Only)**

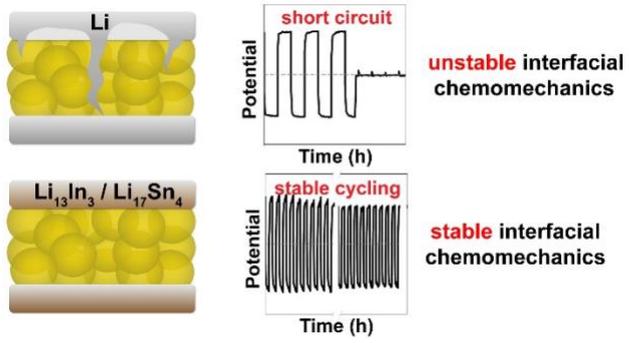